\documentclass[sigconf]{acmart}

\AtBeginDocument{%
  }

\setcopyright{acmlicensed}
\copyrightyear{2026}
\acmYear{2026}
\acmDOI{}

\setcopyright{none} 
\acmConference[BoatSE '26]{7th International Workshop on
Bots and Agents in Software Engineering}{April 13, 2026}{Rio de Janeiro, Brazil}
\acmISBN{}

\usepackage{amsmath}
\usepackage{algorithmic}
\usepackage{graphicx}
\usepackage{booktabs}
\usepackage{multirow}
\usepackage{array}
\usepackage{float}
\usepackage{pifont}
\usepackage{listings}
\usepackage{balance}
\usepackage{cmap}          
\usepackage{threeparttable}
\usepackage{colortbl}
\usepackage[inline]{enumitem}
\usepackage{tikz}
\usepackage[table]{xcolor}
\usepackage{tabularx,array}
\usepackage[most]{tcolorbox}
\usepackage{subcaption}    

\definecolor{paleblue}{rgb}{0.85,0.9,1.0}
\definecolor{deepblue}{rgb}{0.05,0.35,0.75}

\tcbset{
  mybluebox/.style={
    enhanced,
    colback=paleblue,
    frame hidden,
    borderline west={2.5pt}{0pt}{deepblue},
    boxsep=1ex,
    sharp corners
  }
}

\newcolumntype{Y}{>{\raggedright\arraybackslash}X}

\newcommand{\circdiam}{1.3em}
\DeclareRobustCommand{\circnum}[1]{%
  {\footnotesize
   \tikz[baseline=(n.base)]\node (n)
     [circle, fill=black, inner sep=0.8pt, minimum size=\circdiam]
     {\color{white}\sffamily\bfseries #1};}%
}
\newcommand{\circref}[1]{\hyperref[#1]{\circnum{\ref*{#1}}}}

\newcommand{\rqone}{\textbf{RQ1:} How does query reformulation help in localizing files in a large repository?~}
\newcommand{\rqtwo}{\textbf{RQ2:} How does upstream query reformulation improve localization accuracy for LLM-based agents?~}

\begin{document}

\title{Reformulate, Retrieve, Localize: Agents for Repository-Level Bug Localization}

 \author{Genevi\`eve Caumartin}
 \affiliation{
   \institution{Concordia University}
   \city{Montreal}
   \country{Canada}
 }
 \email{genevieve.caumartin@mail.concordia.ca}

 \author{Glaucia Melo}
 \affiliation{
   \institution{Toronto Metropolitan University}
   \city{Toronto}
   \country{Canada}
 }
 \email{glaucia@torontomu.ca}

\begin{abstract}
Bug localization remains a critical yet time-consuming challenge in large-scale software repositories. 
Traditional information retrieval–based bug localization (IRBL) methods rely on unchanged bug descriptions, which often contain noisy information, leading to poor retrieval accuracy. Recent advances in large language models (LLMs) have improved bug localization through query reformulation, yet the effect on agent performance remains unexplored.
In this study, we investigate how an LLM-powered agent can improve file-level bug localization via lightweight query reformulation and summarization. We first employ an open-source, non-fine-tuned LLM to extract key information from bug reports, such as identifiers and code snippets, and reformulate queries pre-retrieval. Our agent then orchestrates BM25 retrieval using these preprocessed queries, automating localization workflow at scale. Using the best-performing query reformulation technique, our agent achieves 35\% better ranking in first-file retrieval than our BM25 baseline and up to +22\% file retrieval performance over SWE-agent.
\end{abstract}


\maketitle

\section{Introduction}
As software systems grow in size and complexity, identifying the precise locations of bugs becomes increasingly challenging. Bug localization, the process of identifying the source of software defects, is a critical step in large repositories~\cite{improving_bug_localization,LongCodeArena}. Automated bug localization can help ease the debugging process by focusing on the buggy areas, improving software quality and developer productivity~\cite{flexfl}. Many automated approaches rely on information retrieval (IR) techniques that match keywords from bug reports with those in source code files~\cite{Zhou_Zhang_Lo_2012, Rahman_Roy_2018, Mahmud_2024}. However, when the terminology used in the report differs from that in the code, retrieval accuracy drops: a problem known as the lexical gap~\cite{query_expansion_code_search}. 

Most existing works on information retrieval-based bug localization (IRBL) use unchanged bug descriptions to locate relevant files~\cite{Rahman_Roy_2018}. Consequently, the quality of the information in a bug, such as the inclusion of observed behaviour, steps to reproduce, and expected behaviour, as well as localization ``hints'' such as code snippets or identifiers, is paramount in building effective automated retrieval methods~\cite{Mahmud_2024, search_queries_ir}. 

Prior work has explored query reformulation and summarization for bug localization. Query reformulation expands, reduces, or replaces query terms to emphasize relevant information~\cite{auto_summarization, Rahman_Roy_2018}, while summarization produces concise bug reports that retain essential details~\cite{auto_summarization}. LLMs have recently been used to improve retrieval focus in bug localization tasks by reformulating bug descriptions into more precise queries~\cite{Mahmud_2024,agents_code_search}. While prior work has integrated lexical search tools into LLM-based agents for retrieval or search-space reduction \cite{locagent},\cite{Rafi_Kim_Chen_Wang_2025}, the interaction between query reformulation and downstream agent performance remains under-explored. Existing research primarily evaluates each component in isolation, examining either the retrieval quality of reformulated queries or the reasoning ability of agents, without analyzing how reformulated inputs affect an agent’s ability to locate relevant files. Agent-based retrieval has been proposed to mitigate the issues of long context understanding and to enable large codebase traversal~\cite{locagent}. LLM-based agents are used more and more for diverse software engineering (SE) tasks, such as bug localization, code generation and code review~\cite{swe-agent, agent-coder, agent-code-review}. These agents coordinate retrieval and reasoning steps through tool use, providing a scalable alternative to end-to-end LLM reasoning. 

In this study, we leverage agentic query reformulation and summarization for file-level bug localization. First, we prompt an LLM to extract the relevant information from a bug report, asking it to explain the bug (a summary), along with information extracted from the bug description: file paths and names, identifiers (classes, methods), code snippets, stack traces, and error messages. To retrieve relevant files using this reformulated bug description, a simple lexical search (BM25~\cite{bm25}) can narrow down the repository files to only the top-$k$ best candidates. 
The agent has access to the full bug description, the repository name, and tools to extract relevant information and preview file content. Each run begins with the bug report and a BM25 query built from the extracted fields, producing a ranked list of candidate files. The agent selects files to inspect and reasons about their relevance within a multi-step conversation. This setup allows us to measure how pre-retrieval (upstream) query reformulation affects both retrieval accuracy and post-retrieval (downstream) ranking quality in agent-based bug localization.

Unlike prompt engineering or query expansion, which adjust prompt text to guide generation, our method treats reformulation as an intermediate representation task—transforming noisy bug reports into structured, retrieval-ready summaries optimized for repository-scale search. We then assess how LLM-based reformulation improves BM25 retrieval and benefits agentic workflows. Specifically, we ask:

\textbf{\rqone} We prompt a smaller, open-sourced LLM to extract structured information from each bug description using a predefined JSON schema, and compare BM25 retrieval results against a baseline BM25 search using the full, unmodified bug description.

\textbf{\rqtwo} We use open-source, non-fine-tuned models-Qwen2.5-coder-32B (Qwen2.5-32B) and Qwen3-coder-30B (Qwen3-30B), and integrate BM25 as a tool for search-space reduction within an LLM-based agentic pipeline. 

This work is the first to explore the effect of upstream query reformulation on agent-based bug localization. Specifically, we contribute:
\begin{itemize}
    \item Evaluate the impact of query reformulation using information extraction by an LLM on two long-context datasets: Long Code Arena~\cite{LongCodeArena} and SWE-Bench Lite~\cite{swe-bench}. We report improvements of up to 36\% on MAP@1.
    \item Design a lightweight agent workflow, including tools such as BM25 for space reduction and individual file viewing, to assess the retrieval ability of LLMs as agents against lightweight BM25 retrieval with query reformulation.
    \item Evaluate the impact of upstream query reformulation and show improvements of up to 35\% on first-file retrieval. Our agent achieves improvements of up to 22\%.
    \item Publish all artifacts in our replication package~\cite{replication}.
\end{itemize}

\section{Related Work}

Bug localization research has evolved from traditional information-retrieval (IR) techniques to modern, LLM-driven agentic systems. Early approaches focused on enhancing lexical and structural matching between bug reports and source code; subsequent work concentrated on automating query reformulation to improve retrieval accuracy. Most recently, LLM-based and multi-agent frameworks have extended these ideas to repository-level reasoning and scalable localization.

\noindent
\textbf{Retrieval-Based Bug Localization}.
Early studies on file-level bug localization focused on improving information retrieval techniques. Zhou et al. (BugLocator)~\cite{Zhou_Zhang_Lo_2012} study a revised Vector Space Model (rVSM), outperforming VSM~\cite{vsm}, LDA~\cite{lda}, LSI~\cite{lsi}, and SUM~\cite{sum}. Saha et al. (BLUiR)~\cite{improving_bug_localization} incorporates structured code information such as class and method names. Hybrid methods like DNNLOC~\cite{bug_loc_deep_learning} combine rVSM with deep neural networks, achieving 50\% top-1 localization accuracy.

\noindent
\textbf{Query Reformulation.}
Kim et al.~\cite{Kim_Lee_2019} improved retrieval by automatically selecting and augmenting bug report terms, yielding +17\% top-1 and +10\% Mean Average Precision (MAP)@10 gains. Rahman et al.~\cite{Rahman_Roy_2018} proposed BLIZZARD, which classifies bug report quality and reformulates queries accordingly, improving Hit@10 by 56\% and MAP@10 by 19\%. Mahmud~\cite{Mahmud_2024} highlighted the lexical gap between reporters and developers, showing through ChatGPT-based extraction that LLM reformulation can bridge this gap.

\noindent
\textbf{Agent-Based Retrieval.}
Yang et al. introduced SWE-Agent~\cite{swe-agent}, which integrates keyword-based search and file inspection tools to locate and edit relevant code segments, reaching an 18\% resolution rate on SWE-bench Lite with GPT-4-Turbo. Xie et al.~\cite{xie-etal-2025-swe-fixer} developed SWE-Fixer, a lightweight approach comprising two modules: code retrieval (BM25-based) and code editing, using open-source fine-tuned models. They obtain competitive performance on SWE-bench Lite/Verified (22\% and 30.2\% task resolution, respectively). Chen et al. introduced LocAgent~\cite{locagent}, which uses fine-tuned LLM agents and heterogeneous code graphs with BM25 indexing to locate buggy files, outperforming prior IR methods. Similarly, Rafi et al.~\cite{Rafi_Kim_Chen_Wang_2025} proposed LLM4FL, a multi-agent framework combining graph-based navigation and reasoning, achieving an 18.6\% improvement over the baseline.

We use a lightweight, non-fine-tuned LLM-based agent to extract key information for lexical retrieval. We explicitly examine how reformulating bug descriptions upstream affects both lexical retrieval and the agent’s ability to identify relevant files.

\section{Proposed Method} \label{sec:method}
For \textbf{RQ1}, we select BM25~\cite{bm25}, a lightweight lexical search tool that does not require prior training on our datasets and is widely used in other studies~\cite{swe-bench, locagent, LongCodeArena}. We use the Pyserini toolkit with default BM25 settings to first index all code files in each repository~\cite{Lin_etal_SIGIR2021_Pyserini}. As a baseline, we use bug descriptions as input for the BM25 search, without any modifications. Similar to previous works~\cite{Rahman_Roy_2018, kbl_query_reformulation, rafi_enhancing_fl}, we select three top-$k$ values: $\{1, 5, 10\}$ to evaluate ranking quality (MAP) and the presence of at least one relevant file in the top-$k$ results (Hit@K).

Next, we integrate an LLM in the loop: from a bug description, we ask it to extract relevant information into a JSON schema. We select Qwen3-30B for this task, and set the temperature to 0 for stability. 
Our idea is to primarily use localization hints as strong signals~\cite{search_queries_ir} (variable and method identifiers, code snippets, stack traces, error messages), as they are more likely to match code files. Still, those hints may not always be available in bug reports. For this reason, we include an ``explanation'' as a summary of the bug, effectively eliminating any boilerplate characters included in bug report templates. 

\textbf{Schema definition.}\label{sec:schema_def} We design a JSON schema (Table~\ref{tab:bug-json-bm25}) that contains essential information from a bug, such as \textit{Explanation} (summary of the bug), \textit{Paths}, \textit{Filenames}, \textit{Identifiers} (list of identifiers mentioned in the bug, such as class, method and variable names), \textit{Code snippet}, \textit{Stack trace} and \textit{Error message}. We evaluate the impact of using only the extracted information by running a BM25 search on the combined fields of the JSON schema, removing the field names, commas and double quotes. We exemplify this process in Table~\ref{tab:bug-json-bm25}.

\textbf{Ablation study.} We examine how each component of the schema contributes to the accuracy of BM25 by forming the following groups:
\begin{enumerate*}[label=\circnum{\arabic*}, ref=\arabic*]
\item\label{it:all} All schema information;
\item\label{it:explanation} Explanation field only;
\item\label{it:all_code} All code signals, including identifiers, code snippets, stack trace and error message;
\item\label{it:id_snippet} Only code identifiers and code snippets; 
\item \label{it:exp_code} Only explanation, identifiers and code snippets.
\end{enumerate*} 
We create the groups based on these assumptions: for group~\circref{it:explanation}, we aim to test the inclusion of relevant code signals in the summary; for~\circref{it:all_code}, all fields that contain code-related signals are included; for~\circref{it:id_snippet} we exclude stack traces and error messages as they may consist of extra tokens not found in source files; for~\circref{it:exp_code}: in addition to~\circref{it:id_snippet}, we include a summary of the bug to provide any extra signals that did not fit any of our categories. We assume that files and paths are primarily helpful for LLM-based evaluation, as they might give insights into which files to inspect. Thus, we do not include them in our groups for this experiment. 

\noindent
\textbf{Statistical analysis.}
We apply Mann–Whitney U ~\cite{Mann_Whitney_1947} and McNemar ~\cite{McNemar1947} tests to assess significance~\cite{cliff_interpretation}.

\begin{table}[t]
\centering
\footnotesize
\setlength{\tabcolsep}{6pt}
\renewcommand{\arraystretch}{1.15}
\caption{Example of JSON fields extracted from a bug report.}
\label{tab:bug-json-bm25}
\begin{tabularx}{\columnwidth}{@{}l Y@{}}
\toprule
\textbf{Original Bug} &
\#\#\#\#\# Expected result

`pipenv clean` exits successfully. Virtualenv is created.

\#\#\#\#\# Actual result

\$ pipenv clean

Creating a virtualenv for this project… [...]

Traceback (most recent call last):

in do\_clean [...]

IndexError: list index out of range \\
\bottomrule
\toprule
\textbf{JSON Field} & \textbf{Value} \\
\midrule
Explanation &
The bug occurs when running 'pipenv clean' after removing the virtual environment. The function get\_requirement in utils.py tries to parse dependencies from an empty list, leading to an IndexError. \\
Path & --- \\
Filename & --- \\
Identifiers & \texttt{[get\_requirement, IndexError]} \\
Code snippet & req = [r for r in requirements.parse(dep)][0] \\
Stacktrace & Traceback (most recent call last): in do\_clean [...] \\
Error message & IndexError: list index out of range \\
\bottomrule
\toprule
\textbf{BM25 query} &
The bug occurs when running 'pipenv clean' after removing the virtual environment. The function get\_requirement in utils.py tries to parse dependencies from an empty list, leading to an IndexError.  [get\_requirement, IndexError],  Traceback (most recent call last): in do\_clean [...], IndexError: list index out of range\\
\bottomrule
\end{tabularx}

\end{table}

For \textbf{RQ2}, we propose automating bug localization and query reformulation using a lightweight agent-based workflow. Existing frameworks require a particular format for tool calling, which our models do not uniformly support. Thus, we implement our own agentic loop with tool-calling capabilities. 

\begin{figure*}
    \centering
    \includegraphics[width=0.5\linewidth]{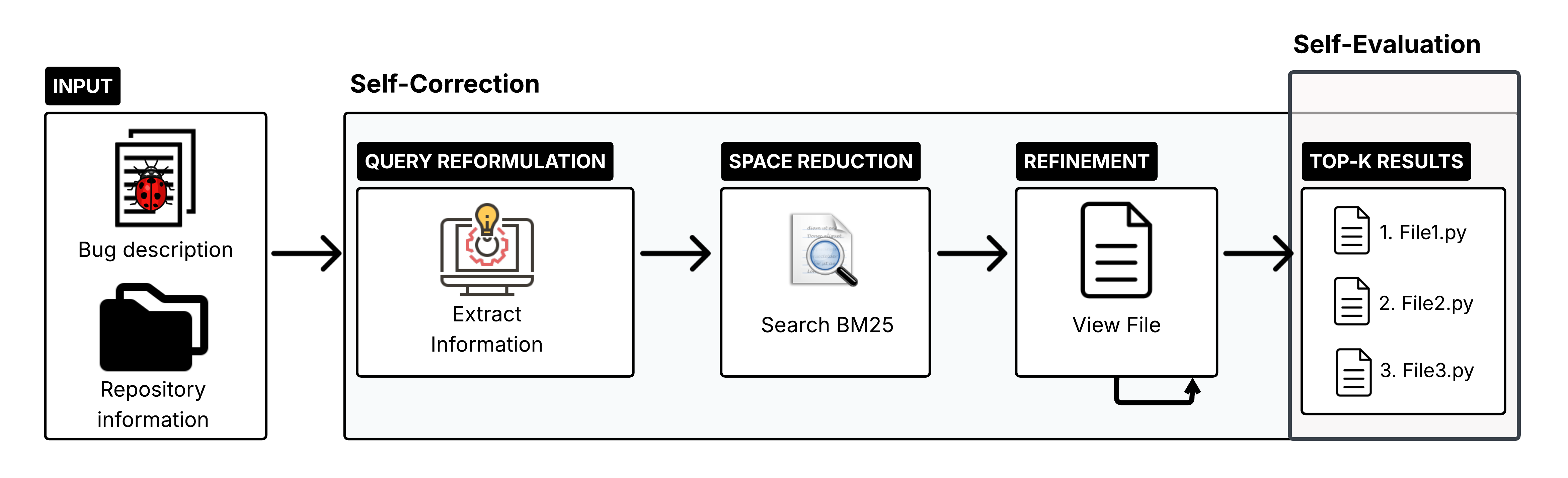}
    \caption{Our Agentic Workflow. At every step, the outputs are validated, and the model is asked to correct invalid outputs.}
    \label{fig:agent_workflow}
\end{figure*}

\noindent
\textbf{Agentic Workflow.} We design an agent that can effectively use available tools to extract the relevant context and localize suspicious files. This agent is guided through the localization process, as shown by our workflow in Figure~\ref{fig:agent_workflow}.
It is divided into steps: 
\begin{enumerate*}
\item Extract relevant information from the bug;
\item Retrieve relevant files using BM25 search;
\item View individual files;
\item Final answer: a ranked list of $n$ candidate files in JSON format. 
\end{enumerate*}
At step 3), the agent is also given the option to exit tool calling and provide its final answer as described in 4). Prompts are provided in our replication package~\cite{replication}.

\noindent
\textbf{Tools.} Our open-sourced models do not support out-of-the-box tool calling. Thus, we design simple functions and instruct the models to return a tool call with parameters in JSON format. We make the following tools available to our agent: 
\begin{enumerate*}
    \item Extract relevant information from a bug description (\textit{extract\_relevant}), returning relevant information from the bug description; 
    \item Retrieve $k$-best results with BM25, with options $k=\{10, 20, 30\}$ (\textit{bm25\_topk}); 
    \item View a particular file given its path (\textit{view\_file});
    \item View the top-level readme file (\textit{view\_readme)}.
\end{enumerate*}

\noindent
\textbf{Input.} We initialize the prompt with the complete bug description and repository name, and instruct the model to retrieve the top-$k$ candidate files ranked by their likelihood of containing the fix ($k = \{1,5,10\}$). During the conversation, the agent has access to the full conversation context: complete bug description, extracted information, and history of tool calls with their results. To manage context size, file views are truncated to 512 tokens. In preliminary experiments, extending views to 1024 tokens yielded only marginal MAP improvements, so we adopt the shorter, more resource-efficient option. 

\noindent
\textbf{Query reformulation.} We design two experiments to assess the benefit of upstream query reformulation for LLMs. 

For our first experiment, the bug description and extracted information always appear at the top of the conversation, and the BM25 query will include all fields; we name this experiment \textit{All-at-top}. With this experiment, we aim to investigate how much the agent can improve on the BM25 results (variation \circref{it:all}), given the same list of candidate files. 

For our second experiment, we replicate a similar setup but restrict the extracted information to the best BM25 result (variation \circref{it:exp_code}). With this setting, BM25 is called with fewer fields.
This experiment aims to demonstrate the improvement our agent achieves using the best-performing BM25 setting. We name this experiment \textit{Best-at-top}.

\noindent
\textbf{Space reduction.} To propose candidate files, the models need to review related files from the project. However, viewing all files is impractical and resource-intensive; context window limitations prohibit processing an entire codebase at once~\cite{locagent}. We apply a space reduction step, building on prior works~\cite{flexfl, swe-bench} prior to LLM-based file localization using BM25. The models are asked to select between three top-$k$ settings for retrieval: \{10, 20, 30\}. 

\noindent
\textbf{Refinement.} Once a list of candidate files is retrieved, the models are given the option to explore individual files. To limit token usage, we provide a partial view of the file, up to 512 tokens, by chunking it from the top. Before chunking, we remove any copyright headers and imports, leaving only the relevant context. Models can view multiple files until they are ready to provide a ranked list of candidate files.

\noindent
\textbf{Self-correction.} Initial observations with our tested models indicate that they occasionally return invalid JSON and incorrect file paths. To mitigate these issues, we implement a simple self-correction mechanism with prompting~\cite{self_correction}: the model's outputs are evaluated for conformity with the JSON format using regular expressions, and file paths are combined with the canonical drive path to confirm their existence. When validation fails, the model is prompted to correct its response given the error. We grant the model three tries to correct its output.

\noindent
\textbf{Self-evaluation.} After a model outputs a response, it is prompted again to validate its answer against all the information it gathered. We start from a fresh context and include information to help the model assess the response: bug description, BM25 results, files viewed and final ranking. The model is asked to review the final ranking and return the list of files with a revised ranking order.

\noindent
\textbf{Baselines.} First, we evaluate the relative accuracy of our agent-based approach against our best BM25 approach from Section~\ref{sec:method}/RQ1 on both LCA and SWE. Then, we evaluate our agentic approach on SWE against prior state-of-the-art approaches: Agentless~\cite{agentless}, LocAgent~\cite{locagent} and SWE-agent~\cite{swe-agent}. SWE-agent does not report file-level localization in their study; thus, we use the results reported by LocAgent. We note that the results reported for Agentless, SWE-agent and LocAgent use Accuracy@K (the fraction of bugs for which the correct file appears within the top-$k$ retrieved results). Because each bug in SWE has a single relevant file, Accuracy@k is equivalent to Hit@K in our setup.

\section{Experiment Settings} \label{sec:experiment_settings}
\subsection{Datasets}
We leverage two datasets that contain real-world SE tasks: SWE-bench-Lite~\cite{swe-bench}, widely used in SE research~\cite{swe-agent, agentless, autocoderover} and Long Code Arena~\cite{LongCodeArena}, whose Bug Localization subset contains 50\% multi-file tasks. 

\begin{table}[!t]
  \centering
  \small
  \begin{threeparttable}
      \caption{Summary statistics for the \textsc{Long Code Arena} and \textsc{SWE-bench-Lite} datasets.}
      \label{tab:dataset-stats}
    
      \begin{tabular}{@{}l c cc cc@{}}
        \toprule
        \textbf{Dataset}
          & \textbf{\# Tasks}
          & \multicolumn{2}{c}{\textbf{Total files}} 
          & \multicolumn{2}{c}{\textbf{Buggy files}}\\
          & &\multicolumn{2}{c}{\textbf{per task}}  & \multicolumn{2}{c}{\textbf{per task}} \\
          \cmidrule(lr){3-4} \cmidrule(lr){5-6}
          &
          & {Mean} & {Max}
          & {Mean} & {Max}\\
        \midrule
        Long Code Arena* & & &\\
         - Bug Localization
          & 150
          & 559 & 7,917
          & 2 & 21\\
        \midrule
        SWE-bench Lite†
          & 300
          & 664 & 895
          & 1 & 1\\
        \bottomrule
      \end{tabular}
      \begin{tablenotes}[flushleft]
      \footnotesize
      \item[*] Source files include only `.py`, `.java` and `.kt` files w/o test files.
      \item[†] Source files include only `.py` files w/o test files.
    \end{tablenotes}
      \end{threeparttable}
\end{table}

\textbf{Long Code Arena (LCA)~\cite{LongCodeArena}} contains six repository-level benchmarks. We focus on the Bug Localization dataset which spans multiple programming languages, including Python, Java, and Kotlin. Our experiments use the test set consisting of 150 manually verified data points — 50 from each language subset. Among these 150 tasks, half involve modifications across two or more files, while the remaining 50\% are single-file tasks.

\textbf{SWE-bench (SWE)~\cite{swe-bench}} tests models' ability to resolve real-world GitHub issues and is exclusively focused on Python. There are three variations of the dataset; we select the SWE-bench Lite variation due to time and resource limitations. This dataset contains 300 tasks that span 11 GitHub repositories, selected from the larger SWE-bench dataset to preserve the distribution and difficulty spectrum of the original benchmark. It contains only single-file tasks.

\subsection{Models}
Many previous works focus on proprietary, larger models~\cite{autocoderover, agentless}; however, results of proprietary models are harder to reproduce, pose serious privacy concerns and may incur significant costs in a typical development pipeline. We pick two models based on our constraints: 1) Ability to run on higher-end consumer hardware; 2) Good instruction following abilities (for structured outputs); and 3) coding abilities.

\noindent
\textbf{Qwen2.5-32B}: Code-specific model in its 32B parameters variant~\cite{qwen2.5-coder}. Qwen2.5-32B was trained on code-specific datasets and scores 83.46\% on the instruction following benchmark IFEval~\cite{ifeval} and 92.1\% on MBPP EvalPlus Base~\footnote{\url{https://evalplus.github.io/leaderboard.html}}.  

\noindent
\textbf{Qwen3-30B}: Code-specific model in its 30.5B parameters variant~\cite{yang2025qwen3technicalreport}. This model has a Mixture-of-Experts (MoE) architecture, where only 3.3B parameters are activated at a given time. It scores 51.6\% on SWE-Bench Verified~\cite{swe-bench} when combined with the OpenHands~\cite{wang2025openhandsopenplatformai} scaffold. As it is a relatively new model at the time of writing, results on other benchmarks are not yet available. 

\noindent
\textbf{Environment.} We use Ollama, version 0.11.10, to host our models locally. All experiments are run on a Mac Studio M4 Max with 64GB of RAM. We limit all models to a 16k context window due to resource constraints, and use greedy decoding (temperature=0) and a fixed seed to minimize variability. However, LLMs are inherently non-deterministic, even in this setting~\cite{song2024goodbadgreedyevaluation}. To mitigate this, we repeat each run three times and report the average. To minimize the variability introduced by LLM-based information extraction during query reformulation, we reuse the best results from Section~\ref{sec:method}/RQ1 in our agent workflow.

\subsection{Metrics}
We use two widely known metrics from information retrieval: Mean Average Precision (MAP) and Hit@K~\cite{flexfl, LongCodeArena, irfl_best_practices}. 
\textbf{MAP} represents the quality of the ranking in retrieval tasks. It measures the average position of all relevant files found by retrieval~\cite{irfl_best_practices, flexfl}. 
\textbf{Hit@K} represents the proportion of rankings that contain at least one valid candidate~\cite{irfl_best_practices}.
Both metrics are complementary; the former evaluates the ranking quality (i.e. order of retrieved files), the latter informs about the proportion of tasks that include at least one relevant file. 

\section{Results}
Table~\ref{tab:results} shows the results of \textbf{RQ1} for both datasets. The first row shows the baseline, which includes the full bug description. The next rows show results for each group, defined in Section~\ref{sec:schema_def}. Indicators show significant differences with the baseline (\ensuremath{\dagger}) and the \textit{explanation-only} (\ensuremath{\ddagger}) variation.

\begin{table}
\centering
\caption{BM25 Baseline vs Extracted Information.} 
\newcommand{\winner}{\cellcolor{gray!25}}
\newcommand{\vsbase}{\textsuperscript{\ensuremath{\dagger}}}
\newcommand{\vsexpl}{\textsuperscript{\ensuremath{\ddagger}}}

  \setlength{\tabcolsep}{1pt} 
  \small
  \begin{threeparttable}
  \begin{tabular}{@{}l|l|lll|lll@{}}
    \toprule
    \multirow{2}{*}{\textbf{Data}} & \multirow{2}{*}{\textbf{BM25 Query}}
      & \multicolumn{3}{c|}{\textbf{MAP}} 
      & \multicolumn{3}{c}{\textbf{Hit@k}} \\
    & & \multicolumn{1}{c}{\textbf{1}} & \multicolumn{1}{c}{\textbf{5}} & \multicolumn{1}{c}{\textbf{10}} & \multicolumn{1}{c}{\textbf{1}} & \multicolumn{1}{c}{\textbf{5}} & \multicolumn{1}{c}{\textbf{10}} \\
    \midrule
\multirow{1}{*}{LCA} & Baseline
  & 0.158 & 0.250 & 0.266
  & 0.320 & 0.653 & 0.713 \\
\midrule
& \circref{it:all} Full
  & 0.229\vsbase\vsexpl & 0.337\vsbase & 0.357\vsbase
  & \textbf{0.447}\vsbase\vsexpl & 0.720 & 0.807\vsbase \\
& \circref{it:explanation} Explanation
  & 0.154 & 0.267 & 0.288
  & 0.300 & 0.667 & 0.767 \\
& \circref{it:all_code} All code
  & 0.233\vsbase\vsexpl & 0.318 & 0.332
  & 0.440\vsbase\vsexpl & 0.653 & 0.720 \\
& \circref{it:id_snippet} Id + snippets
  & \textbf{0.247}\vsbase\vsexpl & 0.336 & 0.349
  & \textbf{0.447}\vsbase\vsexpl & 0.640 & 0.687 \\
& \circref{it:exp_code} 2 and 4
  & \cellcolor{gray!25}0.230\vsbase\vsexpl
  & \cellcolor{gray!25}\textbf{0.339}\vsbase
  & \cellcolor{gray!25}\textbf{0.359}\vsbase
  & \cellcolor{gray!25}0.440\vsbase\vsexpl
  & \cellcolor{gray!25}\textbf{0.733}\vsbase\vsexpl
  & \cellcolor{gray!25}\textbf{0.813}\vsbase\vsexpl \\
      
    \midrule
    \midrule
\multirow{1}{*}{SWE} & Baseline
  & 0.400 & 0.488 & 0.503
  & 0.400 & 0.640 & 0.753 \\
\midrule
& \circref{it:all} Full Schema
  & 0.473\vsexpl & 0.576\vsbase\vsexpl & 0.592\vsbase\vsexpl
  & 0.473\vsbase\vsexpl & 0.740\vsbase\vsexpl & \textbf{0.853}\vsbase\vsexpl \\
& \circref{it:explanation} Explanation
  & 0.370 & 0.434 & 0.446
  & 0.370 & 0.653 & 0.753 \\
& \circref{it:all_code} All code
  & 0.473\vsexpl & 0.554\vsexpl & 0.565\vsexpl
  & 0.473\vsbase\vsexpl & 0.673 & 0.753 \\
& \circref{it:id_snippet} Id + snippets
  & 0.470\vsexpl & 0.538\vsexpl & 0.551\vsexpl
  & 0.470\vsbase\vsexpl & 0.640 & 0.733 \\
& \circref{it:exp_code} 2 and 4
  & \cellcolor{gray!25}\textbf{0.476}\vsexpl
  & \cellcolor{gray!25}\textbf{0.584}\vsbase\vsexpl
  & \cellcolor{gray!25}\textbf{0.597}\vsbase\vsexpl
  & \cellcolor{gray!25}\textbf{0.476}\vsbase\vsexpl
  & \cellcolor{gray!25}\textbf{0.757}\vsbase\vsexpl
  & \cellcolor{gray!25}\textbf{0.853}\vsbase\vsexpl \\
\bottomrule

  \end{tabular}
  \begin{tablenotes}
      \footnotesize
      \item Bold = highest mean in column, shaded = best configuration. 
      \item Superscripts indicate significant improvement at p$<$0.05: † over Baseline, ‡ over Explanation. 
  \end{tablenotes}

  \end{threeparttable}

\label{tab:results}
\end{table}

\textbf{Performance on LCA}. All schema-based variations outperform the BM25 baseline except when using \textit{explanation} only (\circref{it:explanation}) for top-1 retrieval. The \textit{full schema} (\circref{it:all}) performs well overall but is surpassed by \textit{identifiers + snippets} (\circref{it:id_snippet}) and \textit{explanation + identifiers + snippets} (\circref{it:exp_code}) on most metrics except Hit@1. While explanations alone do not provide enough lexical overlap with code, \textit{all code} (\circref{it:all_code}) adds noise, reducing performance at higher $k$. The best overall results are obtained with \textit{explanation + identifiers + snippets} (\circref{it:exp_code}), achieving +31\% MAP@1, +11\% Hit@5, and +12\% Hit@10. For smaller $k$, \textit{identifiers + snippets} (\circref{it:id_snippet}) achieves the highest MAP@1 (+36\%) and Hit@1 (+28\%). In terms of statistical significance, among top schema variants,~\circref{it:all} and~\circref{it:exp_code} are not significantly different at Hit@5/10 and are tied with other variants, except~\circref{it:explanation} at $k$=1. Effect size for significant differences between the best performing variation~\circref{it:exp_code} and the baseline and explanation variations is small on MAP@1, and negligible on Hit@K. 

\textbf{Performance on SWE}. Overall ranking quality and Hit@K scores are higher than on LCA-baseline. MAP@1 is 60\% higher and Hit@1 is 20\% higher, likely due to richer, better-aligned bug reports. As with LCA, all schema-based variations except \textit{explanation} ~\circref{it:explanation} outperform the baseline, achieving up to +16\% MAP and +12\% Hit@10. Differences among schema variations are minor for Hit@K: \textit{explanation} (\circref{it:explanation}), \textit{all code} (\circref{it:all_code}) and \textit{identifiers + snippets} (\circref{it:id_snippet}) have a Hit@10 equivalent to the baseline, suggesting that the benefits are limited to lower $k$ with these variations. Overall,~\circref{it:exp_code} performs best in terms of ranking quality and finding relevant files, although it is equivalent to~\circref{it:all} on Hit@10. In terms of statistical significance, the top variants are tied on MAP and Hit@1, while~\circref{it:all} and~\circref{it:exp_code} are tied on Hit@5 and Hit@10. The best performing variation~\circref{it:exp_code} is significantly different from the baseline and~\circref{it:explanation} (except on MAP@1) with a small effect size.

\textbf{We show our results for \textbf{RQ2} in Table~\ref{tab:results_rq2}.} As a baseline, we use the best-performing variation (\textit{explanation, identifiers and snippets} (\circref{it:exp_code})) query reformulation technique from RQ1 for both LCA and SWE. For SWE, we also compare our results with three prior techniques: SWE-agent~\cite{swe-agent}, Agentless~\cite{agentless}, and LocAgent~\cite{locagent}. 

\begin{table}
\centering
\caption{File-level localization agent results showing averages over three runs.} 
\newcommand{\winner}{\cellcolor{gray!25}}

\newcommand{\up}{\textsuperscript{\textcolor{green!60!black}{\(\blacktriangle\)}}}
\newcommand{\down}{\textsuperscript{\textcolor{red!60!black}{\(\blacktriangledown\)}}}

  \small
  \setlength{\tabcolsep}{2pt} 
  \begin{threeparttable}
  \begin{tabular}{@{}ll|lll|lll@{}}
    \toprule
    \textbf{Dataset /} & \multirow{2}{*}{\textbf{Model}}
      & \multicolumn{3}{c|}{\textbf{MAP}} 
      & \multicolumn{3}{c}{\textbf{Hit@K}} \\
    \textbf{Setting}
    & & \multicolumn{1}{c}{\textbf{1}} & \multicolumn{1}{c}{\textbf{5}} & \multicolumn{1}{c}{\textbf{10}} & \multicolumn{1}{c}{\textbf{1}} & \multicolumn{1}{c}{\textbf{5}} & \multicolumn{1}{c}{\textbf{10}} \\
    \midrule
    \multirow{1}{*}{LCA} 
    & BM25 + \circref{it:exp_code}
      &  0.230 & 0.339 & 0.359
      &  0.440 & 0.733 & 0.813 \\
      \midrule
      
    \textit{All-at-top} & Qwen2.5-32B
      & {0.317} & {0.453} & {0.474}
      & {0.595} & {0.853} & {\textbf{0.878}} \\
    & Qwen3-30B
      & {0.263} & {\textbf{0.472}} & {\textbf{0.490}}
      & {0.513} & {0.831} & {0.867} \\
      
      \textit{Best-at-top} & Qwen2.5-32B
      & {0.324\up} & {0.449\down} & {0.467\down}
      & {0.600\up} & {\textbf{0.860}\up} & {0.860\down} \\
    & Qwen3-30B
      & {\textbf{0.336}\up} & {0.465\down} & {0.478\down}
      & {\textbf{0.627\up}} & {0.853\up} & {0.867} \\
    \midrule
    \midrule
    \multirow{1}{*}{SWE} 
    & BM25 + \circref{it:exp_code}
      & 0.476 & 0.584 & 0.597
      & 0.476 & 0.757 & 0.853 \\
      \midrule
    \textit{All-at-top} & Qwen2.5-32B
      & {0.648} & {0.746} & {0.753}
      & {0.648} & {{0.888}} & {\textbf{0.928}} \\
    & Qwen3-30B
      & {0.674} & {\textbf{0.784}} & {\textbf{0.790}}
      & {0.674} & {{0.888}} & {0.915} \\
  
   \textit{Best-at-top} & Qwen2.5-32B
      & {0.657\up} & {0.758\up} & {0.761\up}
      & {0.657\up} & {\textbf{0.890\up}} & {0.927\down} \\
    & Qwen3-30B
      & {\textbf{0.727\up}} & {0.779\down} & {0.783\down}
      & {\textbf{0.727\up}} & {0.887\down} & {0.910\down} \\
  
    \midrule
    \textit{SWE-agent}\tnote{†} & GPT-4 & 0.573 & --- & --- & 0.573 & 0.690 & --- \\
\textit{Agentless}\tnote{†} & GPT-4o & 0.697 & --- & --- & 0.697 & 0.745 & --- \\
\textit{LocAgent}\tnote{†‡} & Qwen2.5-32B & {0.759} & --- & --- & {0.759} & {0.927} & --- \\
\bottomrule
\end{tabular}
\begin{tablenotes}[flushleft]\footnotesize
\item[‡] Fine-tuned model.
\item[—] Not reported by the source.
\end{tablenotes}
\end{threeparttable}

\label{tab:results_rq2}
\end{table}

Consistent with RQ1 findings, we observe a significant gap between LCA and SWE, both in terms of ranking quality and Hit@K, with SWE reaching 0.928 Hit@10 against 0.880 for LCA, and MAP@10 reaching 0.790 on SWE while LCA maxes at 0.490, a 38\% difference in ranking quality.

\textbf{Performance on LCA}. Both models outperform the baseline in the \textit{All-at-top} experiment. Qwen2.5-32B achieves up to +28\% MAP@1 and +27\% Hit@1, while Qwen3-30B reaches +28\% MAP@5 and +27\% MAP@10, with Hit@5/10 improving by up to 14\%. We observe that gains decline as $k$ increases. Qwen3-30B underperforms on MAP@1, despite leading at higher $k$ values; we analyzed BM25 tool-call patterns across Top-1, Top-5, and Top-10 retrievals but found no clear explanation for this case. The issue likely stems from how Qwen3-30B performs semantic matching based on available signals, as it does not appear in the \textit{Best-at-top} setting or with Qwen2.5-32B. A detailed BM25 tool call analysis can be found in our replication package~\cite{replication}.

Looking into \textit{Best-at-top}, we observe that it yields further improvements at small $k$, reaching +32\% MAP@1 and +30\% Hit@1, but when compared with \textit{All-at-top}, MAP@5 and MAP@10 drop slightly (–1.5\% amd -2\%, respectively) for Qwen3-30B. Qwen2.5-32B only show marginal drop on MAP@5 (-0.8\%) and increase on Hit@5 (+0.8\%). Both models perform better on MAP@1 and Hit@1/5 but see drops in ranking quality on higher MAP.

\textbf{Performance on SWE.} Both experiments show consistent gains over the baseline, with \textit{All-at-top} reaching a near-perfect Hit@10 of 0.928 (+9.5\%). The largest improvements occur at Top-1, where \textit{Best-at-top} increases MAP@1 and Hit@1 by 35\%. For this dataset, however, \textit{Best-at-top} offers limited additional benefit for larger $k$. Comparing best performances: MAP@5 drops slightly (0.784 vs. 0.779), Hit@5 slightly increases, and Hit@10 is almost equal. In terms of model-specific performance, Qwen2.5-32B performs better overall except on Hit@10 where it sees a marginal drop, while Qwen3-30B only performs better on first file retrieval and sees minor drops on other values of $k$.

\textbf{Variability.} For both datasets, we report low variability across our three runs, with standard deviations between $\pm$0.000 and $\pm0.0075$.

\textbf{Prior works.} Comparing with prior works on SWE, LocAgent attains the highest accuracy, about 4\% above our best model on first-file retrieval and Hit@5-but relies on a fine-tuned Qwen2.5-32B. Our approach surpasses SWE-Agent by 21\% on MAP/Hit@1 and 22\% on Hit@5, and exceeds Agentless, achieving +16\% on Hit@5. Notably, our non-fine-tuned models rival or outperform these larger baselines despite being orders of magnitude smaller.

Our results show that lightweight query reformulation is beneficial to first-file retrieval, highlighting the value of upstream query refinement. However, the effect tends to taper at higher $k$. A possible explanation is that paths/filenames are not included in the \textit{Best-at-top} configuration: they may be useful to the agent to better rank files. We leave it for future work to evaluate their impact.

\section{Discussion}

\textbf{Integration in Development Workflows.} Debugging remains one of the most time-consuming and costly SE activities, accounting for 35–50\% of developers' time~\cite{debugging_mindset}. Localizing files relevant to a bug is one of the first steps in a bug-fixing workflow: our approach proposes to assist with that task. One key advantage of our approach is its usability for real-world software development environments: we rely on smaller, open-source LLMs and a lightweight lexical retriever (BM25), which can be deployed without specialized hardware or costly API access.

The information extraction step is both fast and efficient, completing in an average of 5.125 seconds per task on LCA and 4.445 seconds on SWE with Qwen3-30B. The average localization task execution time on LCA lies between $\approx$23-50 seconds and $\approx$18-78 seconds on SWE with Qwen3-30B, the top-10 variations being the most time-intensive. From a practical use standpoint, these results indicate that query reformulation and retrieval can be performed on the fly within a developer’s workflow. This approach is well-suited for integration into IDE assistants where fast, low-cost retrieval is essential.

\noindent
\textbf{Tool Usage and Error Analysis.} Tool-calling preferences vary across models and datasets. To better understand tool usage frequency, we collect detailed statistics, summarized in Table~\ref{tab:tool-usage-errors}.

\textbf{Tool call averages.}
We note that bm25\_topk and extract\_relevant are invoked multiple times per repository. Models have access to prior conversation turns and may choose to re-invoke a previously available tool, which explains this behaviour.

We also observe a high average number of file-view requests per repository. These include: 1) valid file views, where the same file may be opened repeatedly, and 2) invalid paths. To mitigate 1), we warn the model after duplicate views and block repeated access after five occurrences. To address 2), we apply self-correction (Section~\ref{sec:method}/RQ2). After removing duplicate and erroneous calls, file views drop by up to 95\%, highlighting the models’ difficulty in 1) reproducing file paths accurately and 2) following the instruction to view each file only once.
Calls to the view\_readme tool are rare and have a negligible impact on results. 

\textbf{Errors.}
Self-correction covers tool-call validity (correct tool and parameters), file-path verification, and JSON formatting. When an error occurs, the model is warned and given up to three retries. Any model call exceeding ten minutes (600 sec) is cancelled. Table~\ref{tab:tool-usage-errors} reports the corresponding error counts.
We observe more aborted file views with Qwen2.5-32B, whereas Qwen3-30B exhibits repetitive behavior leading to timeouts. JSON formatting errors are corrected more reliably by Qwen2.5-32B, while Qwen3-30B shows a slightly higher number of cases where it was unable to correct its output. 

\begin{table}[t]
\centering
\footnotesize
\setlength{\tabcolsep}{3pt}
\caption{Average tool calls per repository (top) and error counts (bottom) by dataset and model, computed on one run.}
\label{tab:tool-usage-errors}

\begin{threeparttable}
\begin{tabular}{@{}lcccc@{}}
\toprule
& \multicolumn{2}{c}{\textbf{SWE}} & \multicolumn{2}{c}{\textbf{LCA}} \\
\cmidrule(lr){2-3}\cmidrule(lr){4-5}
 & \textbf{Qwen2} & \textbf{Qwen3} & \textbf{Qwen2} & \textbf{Qwen3} \\
\midrule
\textbf{Tool call averages / repo}         & & & & \\
\midrule
{bm25\_top\_k}       & 6.0  & 5.4  & 6.7  & 7.2  \\
{extract\_relevant}  & 6.0  & 4.8  & 6.0  & 6.0  \\
{view\_file}         & 40.2 & 20.1 & 36.8 & 22.2 \\
view\_file unique    & 2.0  & 1.6  & 5.3  & 4.0  \\
{view\_readme}       & 0.0  & 0.0  & 0.0  & 0.0  \\
\midrule
\midrule
\textbf{Error counts}      & & & & \\
\midrule
{Aborted file views (\%)}  & 12.6 & 6.4 & 10.0 & 3.8 \\
{Model timeouts}           & 0    & 18  & 0   & 11  \\
{Aborted - invalid JSON}   & 2    & 12  & 4   & 9   \\
\bottomrule
\end{tabular}
\begin{tablenotes}
\footnotesize
    \item * Qwen2 = Qwen2.5-32B, Qwen3 = Qwen3-30B
\end{tablenotes}
\end{threeparttable}
\end{table}

\section{Threats to Validity}
\noindent
\textbf{Output variability.} LLMs remain non-deterministic, even at temperature 0. To mitigate this threat, we run agent-based experiments three times and report the average of the runs.

\noindent
\textbf{Bug information extraction.} The quality of extracted information can affect retrieval quality; thus, the model used to extract the information and summarize the bug can significantly impact the results. We use Qwen3-30B due to resource constraints.

\noindent
\textbf{Metrics.} This threat concerns whether our chosen metrics accurately capture retrieval performance for bug localization. We use widely accepted metrics (MAP and Hit@K), commonly used in previous studies. 

\noindent
\textbf{Dataset contamination.}
The LLMs we evaluate are pretrained on web-scale code corpora that may include the projects used in LCA and SWE-bench. Observed gains could stem from memorization rather than effective retrieval, and should be interpreted as an upper bound of real-world performance.

\noindent
\textbf{Generalizability.} We run with two different models on two large datasets spanning three programming languages: Java, Python and Kotlin. We expect that the results may not hold for commercial code, models of different sizes or bug reports with a different structure.

\noindent
\textbf{Repository-Level Tasks.} Although SWE-bench Lite contains single-file tasks, they often require searching the entire repository to locate the relevant file, making the problem comparable in complexity to repository-level localization. 

\section{Conclusion}
This study investigates the impact of query reformulation on agent-based workflows for file-level bug localization. Query reformulation, which adds a summary and code signals (identifiers, snippets), consistently strengthens lexical retrieval: with BM25, it boosts first-file localization by 36\% over our baseline. When an LLM-based agent uses the best combination, we achieve an improvement over our best BM25 baseline of 35\%. Our open-sourced, non-fine-tuned LLM agent further advances file-level localization, outperforming SWE-Agent and Agentless by up to 22\%. Future work will explore method-level code chunking for finer retrieval granularity 
and an IDE plugin to assist in file-level bug localization.

\bibliography{main}   

\balance
\end{document}